# Detailing the stress pattern in the area of central Ionian Islands

Anagnostou V.[1], Papadimitriou E.[1], Karakostas V.[1], Bäck T.[2]

(1) Department of Geophysics, Aristotle University of Thessaloniki, Thessaloniki, Greece, vanagno@geo.auth.gr ritsa@geo.auth.gr vkarak@geo.auth.gr
(2) Nuclear Science and Engineering department, KTH Royal Institute of Technology, Stockholm back@kth.se

### Background

The Kefalonia Transform Fault Zone (KTFZ) is the most seismically active area in the Mediterranean and consists of two major branches, the Lefkada fault segment to the north and the Kefalonia fault segment to the south. KTFZ acts as an active boundary between the subduction zone of the remnants of the oceanic lithosphere of the Eastern Mediterranean that subducts under the Aegean microplate to the south and the continental collision between the Eurasian plate and the Adriatic Microplate to the north. The tectonic activity in the region is reflected in the rapid crustal deformation rates of the region and subsequently the frequent occurrence of strong earthquakes ($M_w \geq 6.0$) that occurred during both the historical and instrumental era of seismology. Those strong earthquakes and their temporal distribution can be explained due to stress transfer between closely located fault segments (Papadimitriou, 2002) and as such, studying those stress interactions is an integral part of understanding the long-term tectonic loading in the region.

For better understanding of the kinematics and the distribution of continental deformation that is occurring within the brittle crustal layer in the study area, we assess the regional stress field by inverting focal mechanisms. Contemporary stress field in an area is a result of processes that operate at significantly different spatial scales (Zoback, 1992). Forces that are linked to major tectonic movements such as ridge push and slab pull affect the stress field at large wavelengths, beyond 2000 km and are considered first–order sources of stress field perturbation. At intermediate wavelengths (between 2000 and 500 km) the crustal stress field is affected by processes such as orogenic lithospheric flexure (Zoback, 1992) with those processes representing second order sources of stress field disturbances. Third order patterns of crustal stress field are linked to small–scale variations (<100km) that are produced by processes such as active faulting and at times these variations may lead to the nucleation of potentially destructive earthquakes (Hergert and Heidbach, 2011). Detailed knowledge of the contemporary stress field characteristics of high–risk areas, such the central Ionian Islands region is valuable for possible investigations of stress triggering sensitivity due to the coseismic slip of strong earthquakes.

### Objectives

The main objective of this paper is to present crustal stress data for the area of central Ionian Islands, inferred from the inversion of 485 focal mechanisms using the SATSI inversion method of Hardebeck and Michael (2006) and specifically the MSATSI implementation of the method (Martinez–Garzon et al., 2014). We seek to better detail the stress field by binning the focal mechanism dataset according to known structures in the area and producing a representative focal mechanism for each subarea based on the stress inversion results. Previous work in defining the stress pattern has been done for the Greek region (e.g., Kapetanidis & Kassaras, 2019) including our study area. We focus on detailing the stress pattern around the main branches of the KTFZ along with the transfer zone that connects them and has been identified as an extensional stepover zone (Karakostas et al., 2015).

### Data and Method

Our dataset contains 485 focal mechanisms and was compiled from data obtained from a variety of sources, mainly published works (Hatzfeld et al., 1995; Louvari et al., 1999; Benetatos et al., 2004, 2005; Pondrelli et al., 2004, 2007, 2011; Karakostas & Papadimitriou, 2010; Karakostas et al., 2015; Papadimitriou et al., 2017; Kostoglou et al., 2020) along with routine solutions from various observatories (GCMT, ETH, HUSN, GEOFON). In cases of duplicate solutions, we prioritized the focal mechanism computed by published works over routine solutions. MSATSI deals with the ambiguity of the fault-plane solution by allowing the user to explicitly specify the probability that each input fault plane solution is thought to be the real one. We used the typical 50% value for our inversions. Focal mechanisms that correspond to seismicity occurring below crustal depths (>30km) were discarded in order to constrain the stress patterns to crustal, third order levels. The KTFZ comprises five major fault segments with striking from 12º to 40º, with fault lengths between 12–40km and rake values that exhibit right–lateral strike–slip faulting (Kourouklas et al., 2023). Four out of the five segments bound the western coastlines of the Lefkada and Kefalonia Islands and were identified as the causative faults of four of the strongest earthquakes that occurred in the study area during the 21st century (Karakostas et al., 2004, 2015; Papadimitriou et al., 2017). The fifth segment is located offshore, south of Kefalonia and was associated with the 1983 $M_w$=7.0 earthquake (Scordilis et al., 1985). The gridding scheme for our focal mechanism dataset follows the segmentation of the KTFZ for its five segments, with an additional grid point located in the stepover zone that connects the northern Kefalonia segment with the southern Lefkada segment (Figure 1).

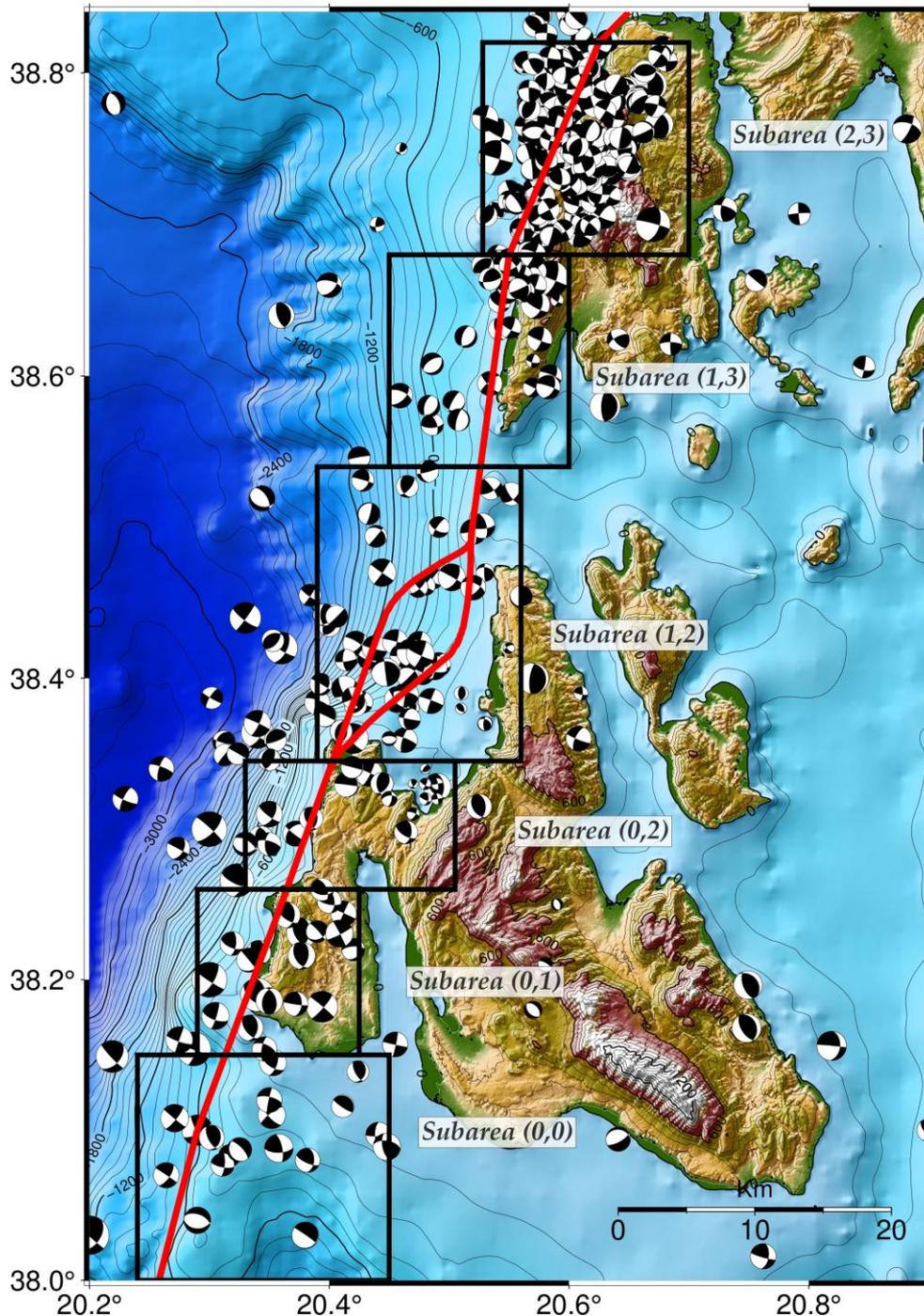

**Figure 1: Map of the study area of the central Ionian Islands along with the major fault segments of the KTFZ and the extensional stepover area, denoted with red lines and enclosed by rectangles that correspond to the subareas or grid points containing the focal mechanisms used in the inversion. The focal mechanisms are shown as equal area lower hemisphere projections.**

After compiling and subdividing our dataset to their corresponding subareas, securing that each one comprises at least 20 focal mechanisms to ensure robustness of results, we perform stress inversion for each subset, by applying the SATSI inversion algorithm. The algorithm inverts focal mechanism data for spatially and/or temporally variable stress field, using a damped least–squares inversion method to obtain a smoothed solution (stress tensor) for each subarea while avoiding data singularities. MSATSI provides the orientation of the three principal axes, as well as a relative stress magnitude quantity (R):

$$R = \frac{\sigma_1 - \sigma_2}{\sigma_1 - \sigma_3} \quad (1)$$

where $\sigma_1$, $\sigma_2$, $\sigma_3$ are the magnitudes of the principal stress axes obtained from the deviatoric stress tensor. This way, the R value quantifies whether the magnitude of the intermediate principal stress $\sigma_2$ is closer to the magnitude of the most compressive or the least compressive principal stress. MSATSI calculates the optimal damping parameter by performing a series of inversions with varying values and comparing the resulting data misfit to the degree of heterogeneity of the solution (model length). The chosen damping parameter is found on the trade–off curve of the two values, corresponding to the point that minimizes both the data misfit and the model length. Finally, we assess the uncertainties of our results using a bootstrap resampling method. We perform resamplings at least 20 times the number of our data for each grid point, specifying the confidence interval of the uncertainties at 95%. Finally, we construct a representative focal mechanism for each grid point, according to the best stress tensor derived from our methodology.

**Exploitable Results**

The results of the damped stress inversion are presented in Figure 2 and Table 1. We observe that in all grid points, the plunge and azimuth of the principal stress axes agree as expected with a strike–slip faulting regime, with low plunge values of both the $\sigma_1$ and $\sigma_3$ principal stress components. Our data constrains the plunge and azimuth of the principal stress axes adequately for the Kefalonia branch of the KTFZ, the stepover zone north of Kefalonia and the southern Lefkada grid point. In the case of the northern Lefkada (grid point 2, 3) subarea, we observe higher uncertainties in the plunges of both the $\sigma_3$ and the $\sigma_2$ principal stress components. Assessing these uncertainties through bootstrap resamplings, we find that they are mainly concerned with the plunges of the $\sigma_3$ axis. This, along with the value of the relative stress ratio R, indicates a stronger normal faulting component in this subarea and thus extensional characteristics. R values for the southern Lefkada (grid point 1, 3) and the stepover zone (grid point 1, 2) are found to be 0.45 and 0.41, respectively. Values of R closer to 0.0 evidence extensional component of the prevailing regional stress field, whereas values approaching 1.0 are indicative of rather transpressional regimes. This is present in the north Kefalonia (grid point 0, 2) south Kefalonia (grid point 0, 1) and offshore Kefalonia (0,0) subareas, where the relative stress magnitude values increase, evidencing a stronger compressive stress component as we move towards the southern edge of the KTFZ. Nevertheless, the R values for all subareas under study remain relatively close to the 0.5 value, meaning the dominant faulting that characterize the regional stress regime is strike–slip. Following the World Stress Map project guidelines (Zoback, 1992) we calculate the azimuth of the maximum horizontal stress axis for every grid point. We observe as expected, that the maximum compression axis $\sigma_1$ corresponds to the $S_{Hmax}$ and strikes SW–NE. Using the best calculated solutions for the P and T axes geometrical characteristics, we produced six representative focal mechanisms, one for each grid point. All are pointing towards a strike–slip stress regime, with the one nodal plane striking NNE–SSW and the second (conjugate) WNW–ESE.

| Grid point (X,Y) | Optimal $\sigma_1$ Trend and Plunge | Optimal $\sigma_2$ Trend and Plunge | Optimal $\sigma_3$ Trend and Plunge | Relative Stress Magnitude | Representative Focal Mechanism ||||||
|---|---|---|---|---|---|---|---|---|---|---|
| | | | | | Nodal Plane 1 ||| Nodal Plane 2 |||
| | | | | | Strike | Dip | Rake | Strike | Dip | Rake |
| (0,0) | 249°/10° | 130°/62° | 341°/11° | 0.76 | 25° | 75° | 179° | 115° | 89° | 15° |
| (0,1) | 248°/11° | 132°/63° | 343°/23° | 0.75 | 21° | 68° | 171° | 114° | 82° | 21° |
| (0,2) | 246°/11° | 123°/70° | 340°/16° | 0.71 | 22° | 73° | 176° | 112° | 86° | 17° |
| (1,2) | 244°/5° | 99°/78° | 337°/6° | 0.41 | 20° | 80° | -177° | 290° | 87° | -9° |
| (1,3) | 242°/13° | 82°/84° | 335°/1° | 0.45 | 19° | 86° | -177° | 289° | 87° | -4° |
| (2,3) | 248°/13° | 10°/67° | 152°/18° | 0.47 | 290° | 67° | 4° | 198° | 86° | 157° |

Table 1: Stress inversion results and representative focal mechanisms corresponding to the optimal principal stress axes for each grid point of our study area.

**Summary and Conclusions**

An effort has been exerted to detail the third–order stress field patterns that dominate the area of central Ionian Islands, the most seismically active part of the broader Aegean area drawing significant interest for its seismotectonic properties.

Taking advantage of the wealth of data provided by the rigorous scientific work done on the region, we find consistent stress field characteristics, typical of dextral strike – slip stress regime. Relative stress magnitude values for each subarea, point towards a notable extensional component to the northern parts of our study area, corresponding to the Lefkada KTFZ fault segments and the extensional duplex zone that connects the southern and northern KTFZ branches, whereas evidencing a transition towards a more transpressional regime towards the Kefalonia branch to the south. Nevertheless, the study area is dominated by the strike–slip fault motions of the KTFZ, indicated by the maximum compression axis being equal to the $SH_{max}$ while the intermediate stress axis plunging steeply in every grid point.

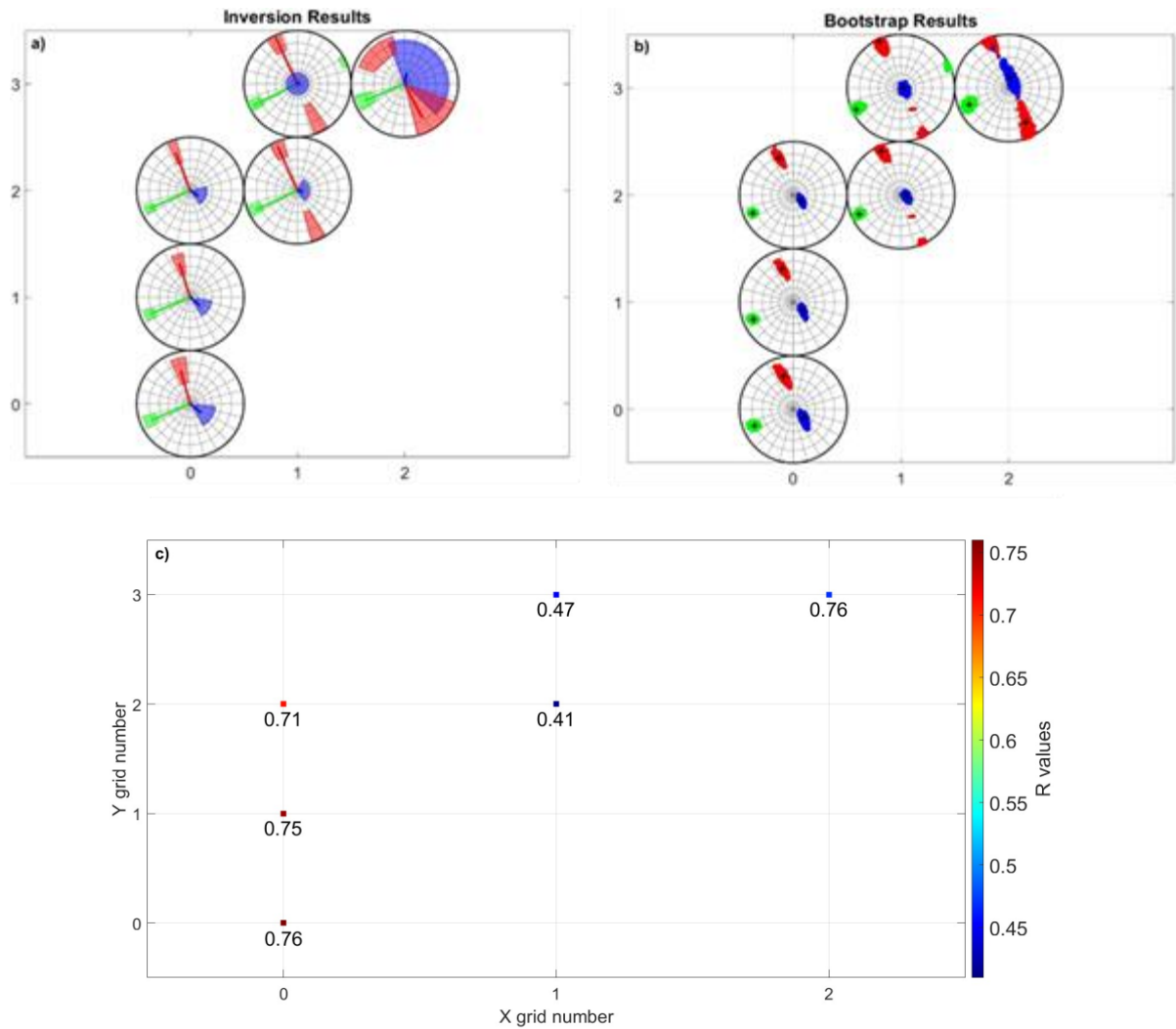

Figure 2: a) Inversion results for the 6 grid points of our study area. Optimal solutions for trend and plunge of the principal stress components $\sigma_1$, $\sigma_2$, $\sigma_3$ (colored green, blue and red, respectively) plotted on Wulff stereographic projection plots, each corresponding to a numbered subarea, according to Table 1. The respective colored areas denote the axes uncertainties in plunge and trend. b) Same as a, but presenting the trend and plunge scatter of the principal stress axes, after 2000 bootstrap resamplings for 95% confidence interval. Black crosses denote the optimal solution for each axis. c) Relative Stress Magnitudes for each of the 6 grid points

### Acknowledgements

This research is financially supported by the artEmis Project funded by the European Union, under Grant Agreement nr 101061712. Views and opinions expressed are however those of the author(s) only and do not necessarily reflect those of the European Union or European Commission – Euratom. Neither the European Union nor the granting authority can be held responsible for them.